\documentclass[12pt]{article}
\usepackage{graphicx}

\newcommand{\imp}{\mbox{\boldmath $p$}}
\newcommand{\kbf}{\mbox{\boldmath $k$}}

\newcommand{\cent}{\centerline}

\newcommand{\orm}{{\rm o}}
\newcommand{\e}{{\rm e}}

\newcommand{\ug}{\; = \;}

\newcommand{\text}{\rm}

\newcommand{\drm}{{\rm d}}

\newcommand{\ze}{\zeta}

\newcommand{\bb}{\begin{equation}}
\newcommand{\ee}{\end{equation}}
\newcommand{\bega}{\begin{eqnarray}}
\newcommand{\ega}{\end{eqnarray}}
\newcommand{\begae}{\begin{eqnarray*}}
\newcommand{\egae}{\end{eqnarray*}}

\newcommand{\h}{\hspace*{4ex}}
\newcommand{\dis}{\displaystyle}

\newcommand{\Om}{\Omega}
\newcommand{\om}{\omega}

\begin{document}

\centerline{Erasmo Recami}

\vspace*{0.2 cm}

\centerline{{\em Facolt\`{a} di Ingegneria, Universit\`{a} statale di Bergamo,
Dalmine (BG), Italy;}}
\centerline{{\em {\rm and} INFN---Sezione di Milano, Milan,
Italy.}}

\begin{center}
{\large {\bf A homage to E.C.G.Sudarshan: Superluminal objects and waves (An updated
overview of the relevant experiments)$^{\: (\dag)}$}}
\footnote{This work has been partially supported by the University of Texas at
Austin, USA, and by the Italian INFN and MIUR. \ E-mail for contacts:  recami@mi.infn.it}
\end{center}

\vspace*{5mm}

{\bf 1. - Introduction.}\hfill\break

\h It is a great privilege and pleasure to be able to present (after a
brief theoretical introduction) a panoramic view of the experiments
that ---revealing the apparent existence of Superluminal group-velocities---
seem to confirm the pioneering works published by E.C.George Sudarshan,
already in the sixties, about tachyons.

The question of Superluminal ($V^{2}>c^{2}$) objects or
waves has a long story, starting perhaps in 50 B.C. with Lucretius'
{\em De Rerum
Natura}\footnote{Cf., e.g., book 4, line 201: ``Quone vides {\em citius} debere
et longius ire / Multiplexque loci spatium transcurrere eodem / Tempore {\em
quo Solis} pervolgant {\em lumina} coelum?"}. Still in
pre-relativistic times, one meets various related works, from those by
J.J.Thomson to the papers by A.Sommerfeld.

With Special Relativity, however, since 1905 the conviction spread over that the
speed $c$
of light in vacuum was the {\em upper} limit of any possible speed.

For instance, R.C.Tolman in 1917 believed to have shown by his ``paradox" that
the existence of particles endowed with speeds larger than $c$ would have
allowed sending information into the past. Such a conviction blocked for
more than half a century -- aside from an isolated paper (1922) by the
Italian mathematician G.Somigliana -- any research about Superluminal
speeds. Our problem started to be tackled again essentially in the fifties
and sixties, in particular after the mentioned, epoch-making papers by
George Sudarshan {\em et al.\/}[1], which provoked much further
work, in particular by E.Recami and coworkers[2], as well as by H.C.Corben
and others (to confine ourselves to the {\em theoretical} researches). The
first experiments looking for faster-than-light objects were performed by
T.Alv\"{a}ger {\em et al.\/}[2].

\h Superluminal objects were called tachyons, T, by G.Feinberg,
from the Greek word $\tau \alpha \chi {\acute{\upsilon}}
\varsigma$, quick (and this induced the present author in 1970 to
coin the term bradyon, for ordinary subluminal ($v^2<c^2$)
objects, from the Greek word $\beta \rho \alpha \delta
{\acute{\upsilon}} \varsigma$, slow).  Finally, objects traveling
exactly at the speed of light are called ``luxons".

\h  In recent years, terms as ``tachyon'' and ``superluminal''
fell unhappily into the (cunning, rather than crazy) hands of
pranotherapists and mere cheats, who started squeezing money out of
simple-minded people; for instance by selling plasters (!) that should cure
various illnesses by ``emitting tachyons''... We are dealing with tachyons
here, however, since at least four different experimental sectors of physics
seem to indicate the actual existence of Superluminal motions
(thus confirming long-standing theoretical predictions [1,3]).

In the first part of this article (after a brief, non-technical theoretical
introduction, which can be useful since it informs about still scarcely known
approaches) we mention the various experimental sectors of physics
in which
Superluminal motions seem to appear. In particular, a bird's-eye view is
presented of the experiments with evanescent waves (and/or tunnelling photons),
and  with the ``localized Superluminal solutions" (SLS) to the wave
equations, like the so-called X-shaped waves; the shortness of this review
is compensated by a number of references, sufficient in some cases to provide
the interested readers with reasonable bibliographical information.

\

\

{\bf 2. - General concepts}\hfill\break

\ As far as classical tachyons are concerned, let us insert Sudarshan's
original contributions within the picture provided by Special Relativity
(SR), once one {\em does not restrict} it[2,3] to subluminal motions.

\h Let us premise that SR, abundantly
confirmed by experience, can be built on the two simple, natural Postulates:

1) that the laws (of electromagnetism and mechanics) are valid not only for a
particular observer, but for the whole class of the ``inertial"
observers;

2) that space and time are homogeneous and space is moreover isotropic.

From these Postulates one can theoretically {\em infer} that one, and only
one, {\em invariant} speed exists: and experience tells us such a speed to
be the one, $c$, of light in vacuum (namely, 299.792458 km/s).
Indeed, ordinary light possesses the peculiar feature of presenting always
the same speed in vacuum, even when we run towards or away from it.
It is just that
feature, of being invariant, that makes the speed $c$ quite exceptional:
no bradyons, and no tachyons, can enjoy the same property.

\h  Another (known) consequence of our Postulates is that the
total energy of an ordinary particle increases when its speed $v$ increases,
tending to infinity when $v$ tends to $c$. Therefore, infinite forces would
be needed for a bradyon to reach the speed $c$. This fact generated the
popular opinion that speed $c$ can be neither achieved nor overcome.

However, as speed $c$ photons exist which are born, live, and die always at
the speed of light[1] (without any need of accelerating from rest to the light
speed), so objects can exist[4] always endowed with speeds
$V$ larger than $c$ (see Fig.1). This circumstance has been picturesquely
illustrated by Sudarshan (1972) with reference to an imaginary
demographer studying the population patterns of the Indian subcontinent:

\begin{quote}\small
Suppose a demographer calmly asserts that there are no people North of the
Himalayas, since none could climb over the mountain ranges! That would be an
absurd conclusion. People of central Asia are born there and live there:
they did not have to be born in India and cross the mountain range. So with
faster-than-light particles.
\end{quote}\normalsize

\begin{figure}[!h]
\begin{center}
 \scalebox{0.6}{\includegraphics{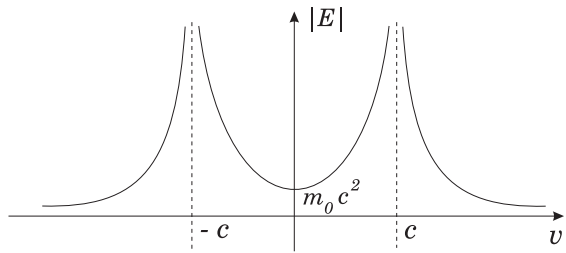}}
\end{center}
\caption{Energy of a free object as a function of its speed.[1-4]} \label{fig1}
\end{figure}

Let us add that, still starting from the
above two Postulates (besides a third postulate, even more
obvious\footnote{Namely, the assumption that there are no particles
---regularly traveling forward in time--- endowed with negative energies.},
the theory of relativity can be generalized[1,3] in such a way as to
accommodate also Superluminal objects; such a {\em non-restricted} version of
SR is sometimes called ``extended relativity".  Also within extended
relativity[3] the
speed $c$, besides being invariant, is a limiting velocity: but every
limiting value has two sides, and one can a priori approach it both from
the left and from the right.

Actually, as we were saying, the ordinary formulation of SR has
been restricted {\em too much}. \ For instance, even leaving
Superluminal speeds aside, it can be easily so widened as to
include antimatter[5]. Then, one finds space-time to be a priori
populated by normal particles P (which travel forward in time
carrying positive energy), {\em and} by dual particles Q ``which
travel backwards in time carrying negative energy''. The latter
shall appear to us as antiparticles, i.e., as particles --
regularly traveling forward in time with positive energy, but --
with all their ``additive'' charges (e.g., the electric charge)
reversed in sign[5,1]: see Fig.2.

\begin{figure}[!h]
\begin{center}
 \scalebox{0.6}{\includegraphics{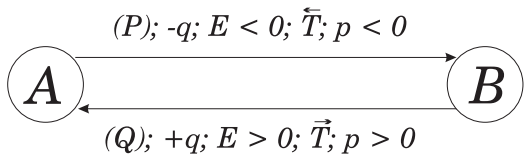}}
\end{center}
\caption{Depicting the ``switching rule" (or reinterpretation principle) by
Stueckelberg-Feynman-Sudarshan[1-5]: \ $Q$ will appear as the
antiparticle of P. \ See the text.} \label{fig2}
\end{figure}

To clarify this point, we can here recall only what follows: We,
as macroscopic observers, have to move in time along a single,
well-defined direction, to such an extent that we cannot even  see
a motion backwards in time... and every object like Q, traveling
backwards in time (with negative energy), will be {\em
necessarily} reinterpreted by us as an anti-object, with opposite
charges but traveling forward in time (with positive energy): cf.
Fig.2 and refs.[3-5,1].

But let us forget about antimatter and go back to ``tachyons". A
strong objection against their existence is based on the opinion
that by using tachyons it would be possible to send signals into
the past, owing to the fact that a tachyon T which, say, appears
to a first observer $O$ as emitted by A and absorbed by B, can
appear to a second observer $O^{\prime}$ as a tachyon T' which
travels backwards in time with negative energy[1,3]. However, by
applying (as it is obligatory to do) the same ``reinterpretation
rule" or switching procedure seen above, T' will appear to the new
observer $O^{\prime}$ just as an antitachyon ${\overline{{\rm
T}}}$ emitted by B and absorbed by A, and therefore traveling
forward in time, even if in the contrary {\em space} direction. In
such a way, every travel towards the past, and every negative
energy, disappear[1,3-5]. \ The mentioned reinterpretation
procedure[1,3-5] ought to be called the Sudarshan's principle, or
the Stueckelberg-Feynman-Sudarshan principle: indeed, it was
Sudarshan[1] who stated it clearly, by taking proper account of
the interplay between the signs {\em both} of the motion direction
in time {\em and} of the energy.

Starting from this observation, it is possible to solve[1,6] the so-called
causal paradoxes associated with Superluminal motions: paradoxes which
result to be the more instructive and amusing, the more sophisticated they
are, but that cannot be re-examined here\footnote{Some of them have been
proposed by R.C.Tolman, J.Bell, F.A.E.Pirani, J.D.Edmonds and
others[1,6,3,2].}

Let us mention here just the following. The reinterpretation principle,
according to which, in simple words, signals are carried only by objects
which appear to be endowed with positive energy, does eliminate any
information transfer backwards in time; but this has a price: that of
abandoning the ingrained conviction that the judgement about what is cause
and what is effect is independent of the observer[1-6]. In fact, in the case
examined above, the first observer $O$ considers the event at A to be the
cause of the event at B. By contrast, the second observer $O^{\prime}$
will consider the event at B as causing the event at A. \ All the observers
will however see the cause to happen {\em before} its effect[1-6].

Taking new objects or entities into consideration always
forces us to a criticism of our prejudices. If we require the phenomena to
obey the {\em law} of (retarded) causality with respect to all the
observers, then we cannot demand also the {\em description} ``details'' of
the phenomena to be invariant: namely, we cannot demand in that case also the
invariance of the ``cause'' and ``effect'' {\em labels\/}[6,2].

To illustrate the nature of our difficulties in accepting that, e.g., the
labels of cause and effect depend on the observer, let us cite an analogous
situation that does not imply present-day prejudices:

\begin{quote}\small
For ancient Egyptians, who knew only the Nile and its tributaries, which all flow South to North,
the meaning of the word ``south'' coincided with the one of ``upstream'', and the meaning of the
word ``north'' coincided with the one of ``downstream''. When
Egyptians discovered the Euphrates, which unfortunately happens to flow
North to South, they passed through such a crisis that it is mentioned in
the stele of Tuthmosis I, which tells us about {\em that inverted water that
goes downstream (i.e. towards the North) in going upstream} [Csonka,
1970]. \end{quote}\normalsize

In the last century, theoretical physics led us in a natural way
to suppose the existence of various types of objects: like magnetic monopoles,
quarks, strings, tachyons, besides black-holes etcetera: and various sectors of
physics could not go on without them, even if the existence of most of them
is uncertain (perhaps, also because attention has not yet been paid to some
links existing among them: e.g., a Superluminal electric charge is expected to
behave as a magnetic monopole; and a black-hole a priori can be the source
of tachyonic matter).  According to Democritus of Abdera, everything that
was thinkable without meeting contradictions had to exist somewhere in the
unlimited universe. This point of view -- which was given by M.Gell-Mann the
name of ``totalitarian principle'' -- was later on expressed (T.H.White) in
the humorous form ``Anything not forbidden is compulsory''...

\

\

\

{\bf 3. A glance at the experimental status-of-the-art}.\hfill\break

\h Extended Relativity can allow a better understanding of many
aspects also of {\em ordinary} physics; and this remains true independently
of the circumstance that tachyons do or do not exist in our cosmos as
asymptotically free objects (their existence as ``intermediate states" is,
of course, obvious[1,3]). As already said,
we are dealing with Superluminal motions, however, since this topic has
recently returned in fashion, especially because of the fact that at least three
or four different experimental sectors of physics seem to suggest the
possible existence of faster-than-light motions. Our first aim is putting
forth in the following some information about the
experimental results obtained in a couple of those different physics sectors,
with a mere mention of the others.

\

{\bf A)} \ {\bf Neutrinos} --- A long series of experiments, started
in 1971, seems to show that the square ${m_{0}}^{2}$ of the mass $m_{0}$ of
muon-neutrinos, and more recently of electron-neutrinos too, is
negative; which, if confirmed, would mean that (when using a na\"{\i}ve
language, commonly adopted) such neutrinos possess an ``imaginary mass'' and
are therefore tachyonic, or mainly tachyonic[7,3]. Notice, incidentally,
that in extended
relativity the dispersion relation for a free Superluminal object becomes

\[ \om^2-\kbf^2=-\Om^2, \; \mbox{or}\; E^2-\imp^2=-m_\orm^{2},\]

and there is  {\em no} need therefore of imaginary masses!\footnote{We put
$c=1$, whenever convenient, throughout this paper.}  The present author can
testify that at least by 1971 ---and probably some years before (as well
as we ourselves, by the way)--- George Sudarshan had got the idea that
neutrinos could be tachyons: an idea proposed in print by Cawley[7] later
on, in 1972.

\

{\bf B)} \ {\bf Galactic Micro-quasars} --- As to the {\em apparent}
Superluminal expansions observed in the core of quasars[8] and, recently, in
the so-called galactic microquasars[9], we shall not deal here with that
problem, because it is far from the other topics of this paper: not to mention
that for those astronomical observations there exist orthodox
interpretations, based on ref.[10], that ---even if ``statistically"
weak--- are accepted by the majority
of the astrophysicists.\footnote{For a theoretical discussion, see ref.[11].}

Here, let us mention only that simple geometrical considerations
in Minkowski space show that a {\em single} Superluminal light
source would appear[11,3]: \ (i) initially, as a source in the
``optical boom'' phase (analogous to the acoustic ``boom''
produced by an airplane traveling with constant supersonic speed):
namely, as an intense source which suddenly comes into view; and
that \ (ii) afterwards seems to split into TWO objects receding
one from the other with speed \ $V>2c$ [both phenomena being
similar to those actually observed, according to refs.[9]].

\

{\bf C)} \ {\bf Evanescent waves and ``tunnelling photons''} ---
Within quantum mechanics (and precisely in the {\em tunnelling} processes),
it had been shown that the tunnelling time ---firstly evaluated as a simple
``phase time'' and later on calculated through the analysis of the
wavepacket behaviour--- does not depend on the barrier width in the case of
opaque barriers (``Hartman effect'')[12]. This implies Superluminal and
arbitrarily large (group) velocities $V$ inside long enough barriers: see
Fig.3.

\begin{figure}[!h]
\begin{center}
 \scalebox{0.8}{\includegraphics{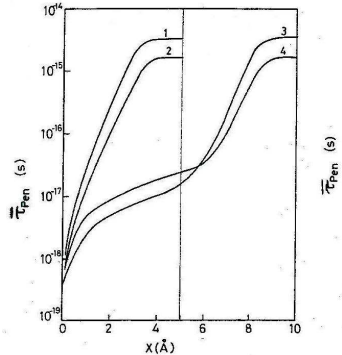}}
\end{center}
\caption{Behaviour of the average ``penetration time" (in seconds) spent by
a tunnelling wavepacket, as a function of the penetration depth (in
{\aa}angstroms) down a potential barrier (from Olkhovsky {\em et al.}, ref.[12]).
According to the predictions of quantum mechanics, the wavepacket speed
inside the barrier increases in an unlimited way for opaque barriers; and the
total tunnelling time does {\em not} depend on the barrier width[12].} \label{fig3}
\end{figure}

Experiments that may verify this prediction by, say, electrons are
difficult: And, in fact, only preliminary results for tunnelling neutrons
exist[12,26]. Luckily enough, however, the Schroedinger equation in the
presence of a potential barrier is mathematically identical to the Helmholtz
equation for an electromagnetic wave propagating, for instance, down a
metallic waveguide along the $x$-axis (as recalled, e.g., by R.Chiao
{\em et al.\/}[13]);
and a barrier height $U$ greater than the electron energy $E$ corresponds
(for a given wave frequency) to a waveguide of transverse size lower than a
cut-off value. A segment of ``undersized" guide ---to go on with our example---
does therefore behave as a barrier for the wave (photonic barrier)[16,13],
as well as any other photonic band-gap filters. The wave assumes therein
---like a particle inside a quantum barrier--- an imaginary momentum or
wave-number and gets, as a consequence, exponentially damped along $x$. In
other words, it
becomes an {\em evanescent} wave (going back to normal propagation, even
if with reduced amplitude, when the narrowing ends and the guide returns to
its initial transverse size). Thus, a tunnelling experiment can be
simulated[13,16] by having recourse to evanescent waves (for which the
concept of group velocity can be properly extended[14]).

The fact that evanescent waves travel with Superluminal speeds (cf., e.g.,
Fig.4) has been actually {\em verified} in a series of famous experiments.

\begin{figure}[!h]
\begin{center}
 \scalebox{1.6}{\includegraphics{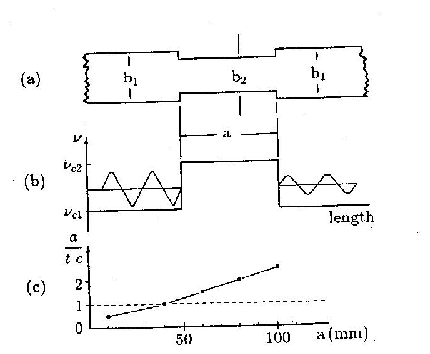}}
\end{center}
\caption{Simulation of quantum tunnelling by experiments with
classical evanescent
waves (see the text), which were predicted to be Superluminal also on the
basis of extended relativity[3,4]. The figure shows one of the measurement
results in refs.[15]; that is, the wavepacket average speed while crossing
the evanescent region ( = segment of undersized waveguide, or ``barrier")
as a function of its length. As theoretically predicted[19,12], such an
average speed exceeds $c$ for long enough ``barriers".} \label{fig4}
\end{figure}

Namely, various experiments, performed since 1992 onwards by G.Nimtz {\em et
al.} at Cologne[15], by R.Chiao, P.G.Kwiat and A.Steinberg's
group at Berkeley[16],
by A.Ranfagni and colleagues at Florence[17], and by others at
Vienna, Orsay, Rennes[17] etc., verified that ``tunnelling photons" travel with
Superluminal group velocities.\footnote{Such experiments raised a great deal of
interest[18], also within the non-specialized press, and were commented on by
{\em Scientific American}, {\em Nature}, {\em New Scientist}, {\em Newsweek},
etc.}
 \ Let us add that also extended relativity had predicted[19] evanescent waves
to be endowed with faster-than-$c$ speeds; the whole matter
appears to be therefore theoretically self-consistent. The debate
in the current literature does not refer to the experimental
results: which can be correctly reproduced even by numerical
elaborations[20,21] based on Maxwell equations only (cf., e.g.,
the illuminating figures 11 and 13 in ref.[21], here reproduced as
Figs.5 and 6 of this paper), but rather to the question whether
they allow, or do not allow, sending signals or information with
Superluminal speed[22,21,14]. \ Actually, a peaked wavepacket
suffers a strong amplitude attenuation while traveling inside a
quantum or classical barrier; its width, however, remains
unaffected (cf. Fig.7): Something that might have some relevance
when thinking of attempting transmissions by Morse's alphabet). \
Moreover, many authors have emphasized that ---at least in the
case of quantum barriers--- the tunnelling of particles is a
statistical process, in the sense that one cannot know a priori
which particle, or photon, will pass through the barrier. This is
true, of course; but the weight of such a consideration becomes
lower when the number of the particles at our disposal for
attempting a ``signal" transmission does increase. To remain
within the Morse alphabet example, one can send out {\em dots} and
{\em dashes} by emitting pulses of, say, one thousand and ten
thousand particles each, respectively; the dot and dashes will
then be recognized also after the tunnelling!... This becomes even
more meaningful when one approaches the classical limit. \  The
claim that superluminal tunnelling cannot be used to transmit any
information is in need, therefore, of further discussion,
especially at the light of what will follow below: More details
can be found, anyway, in refs.[50,51].

\begin{figure}[!h]
\begin{center}
 \scalebox{0.95}{\includegraphics{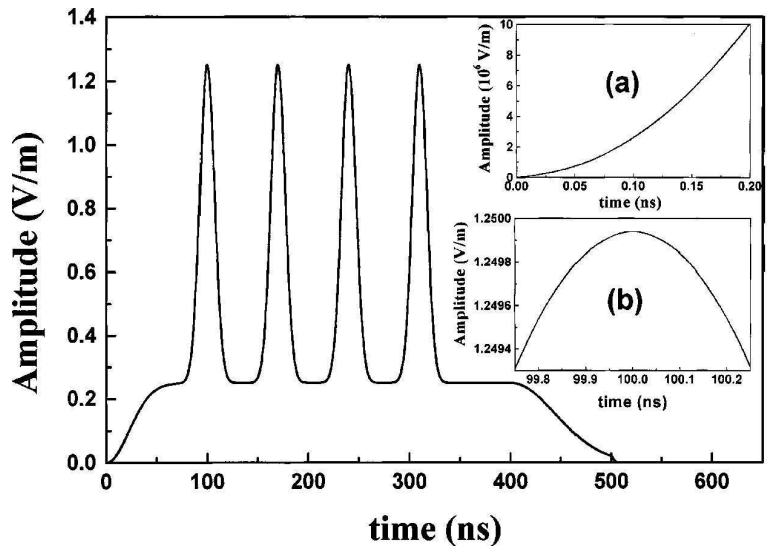}}
\end{center}
\caption{The delay of a wavepacket crossing a barrier (e.g., a classical
barrier: cf. Fig.4) is due to the initial discontinuity: In
ref.[21]
suitable numerical simulations were therefore performed by considering
an (indefinite) {\em undersized} waveguide, and therefore eliminating
any geometric discontinuity in its cross-section.   \
This figure shows the envelope of the initial signal. \ Inset (a) depicts in
detail the initial part of this signal as a function of time, while inset (b)
depicts the gaussian pulse peak centered at $t = 100$ ns.} \label{fig5}
\end{figure}

\

\begin{figure}[!h]
\begin{center}
 \scalebox{0.5}{\includegraphics{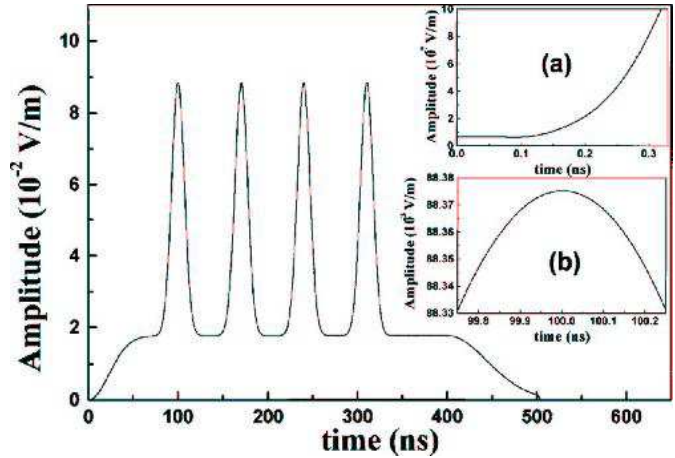}}
\end{center}
\caption{Envelope of the signal in the previous figure
after having traveled a distance $L = 32.96$ mm through the mentioned
undersized waveguide. \ Inset (a) shows in detail the initial
part (in time) of such arriving signal, while inset (b) shows the peak of
the gaussian pulse that had been initially modulated by centering it at $t =
100$ ns. \ One can see that its propagation took {\em zero} time, so that
the signal traveled with infinite speed. \ The numerical simulation has
been based on Maxwell equations only. \ Going on from Fig.5 to this Fig.6
one verifies that the signal strongly lowered its amplitute: However, the
width of each peak did not change (and this might have some relevance
when thinking of a Morse alphabet ``transmission": see Fig.7 and the text.} \label{fig6}
\end{figure}

\

\begin{figure}[!h]
\begin{center}
 \scalebox{0.32}{\includegraphics{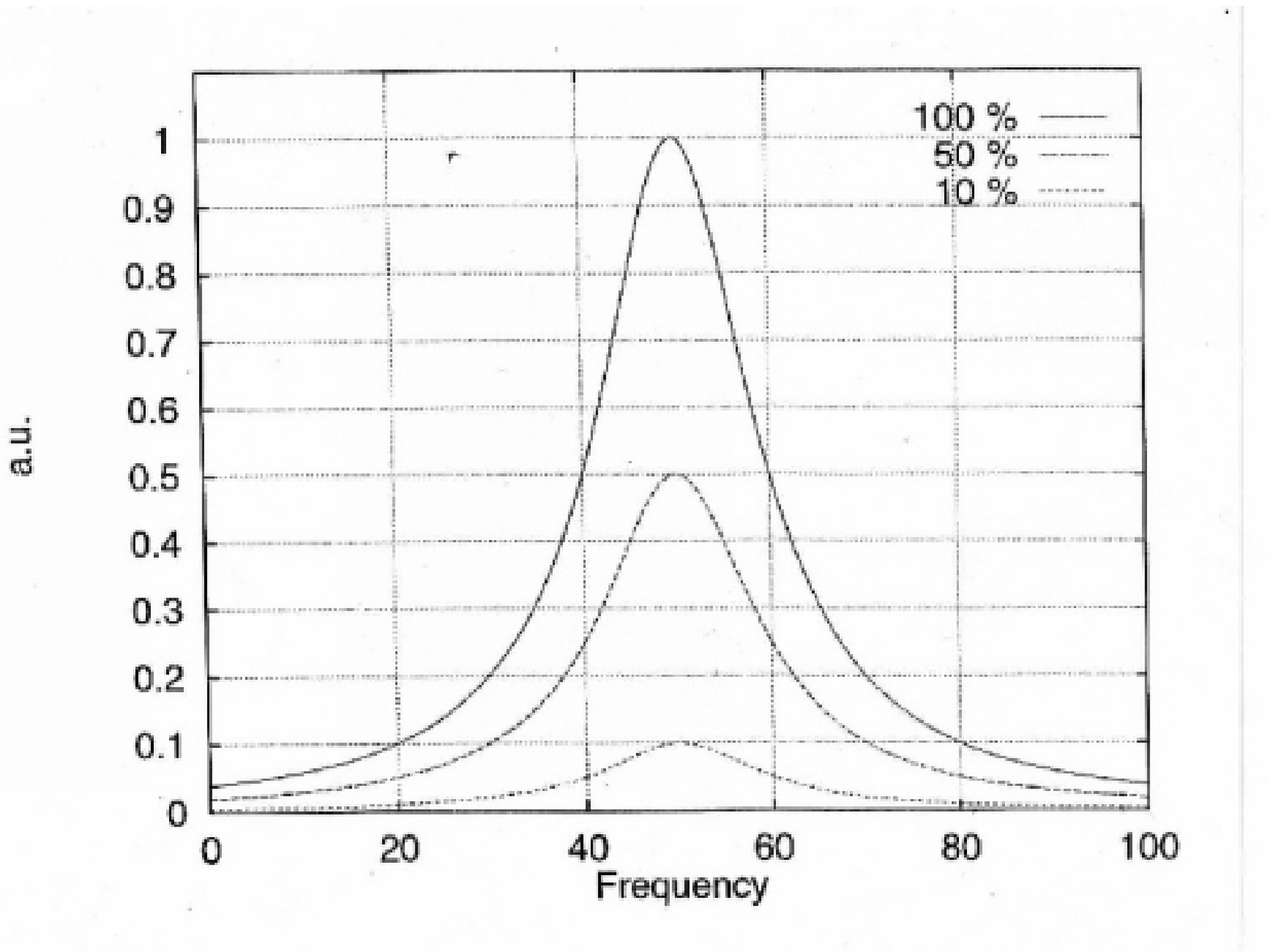}}
\end{center}
\caption{As we saw in Figs.5 and 6, in connection with (classical,
in particular) barriers, a peaked wavepacket suffers a strong
amplitude attenuation while traveling inside a quantum or
classical barrier; its width, however, remains unaffected (this
figure is due to G.Nimtz): Something that, as mentioned in the
previous caption, might have some relevance when thinking of
attempting transmissions by Morse's alphabet.} \label{fig7}
\end{figure}

Let us repeat that all the phenomena mentioned in this, as well as in the
following, sub-sections can be accomodated into the standard frameworks of
quantum physics or of classical relativistic physics: It is therefore
obvious that such phenomena can receive explication or interpretation in
terms of standard physics (e.g., sometimes but not always, in terms of
suitable ``reshapings"): but this does not eliminate the fact that
Superluminal motions take place.

As we already said, in the above-mentioned experiments one meets a substantial
attenuation of the considered pulses during tunnelling (or during propagation
in an absorbing medium). However, by having recourse to suitable devices, as
a ``gain doublet", it has been recently
reported the observation of undistorted pulses propagating with Superluminal
group-velocity with a {\em small} change in amplitude[23].

Let us underline that some of the most interesting experiments of
this series seem to be the ones with TWO ``barriers" (e.g., with two
gratings in an optical fiber, or with two segments of
undersized waveguide separated by a piece of normal-sized waveguide: Fig.8).
For suitable frequency bands ---i.e., for ``tunnelling" far from
resonances---, it was found that the total crossing time does {\em not}
depend on
the length of the intermediate (normal) guide: namely, that the wavepacket
speed along it is infinite[24,25]... This agrees with what predicted by Quantum
Mechanics for the non-resonant tunnelling through two successive opaque
barriers (namely, the tunnelling phase time, which depends on the entering energy,
has been shown by us to be {\em independent} of
the distance between the two barriers[26]); something that has been
theoretically confirmed, and generalized, by Y.Aharonov {\em et al.}[26]. \
Such a prediction has been experimentally verified a second time, with a
cleaner experiment,
taking advantage of the circumstance that quite interesting evanescence
regions can be constructed in the most varied manners, like by means of
different photonic band-gap materials or gratings (it being possible to use
from multilayer dielectric mirrors, or semiconductors, to photonic
crystals...). Indeed, a very recent confirmation came from an experiment
having recourse to two gratings in an optical fiber[25]. On this respect,
rather interesting are the figures 1 and 5 of ref.[25], here reported as
Figs.9 and 10 of this paper (see especially the experimental results
depicted in Fig.10).

\begin{figure}[!h]
\begin{center}
 \scalebox{0.8}{\includegraphics{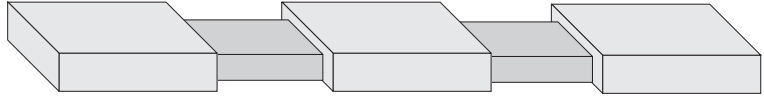}}
\end{center}
\caption{The very interesting experiment along a metallic waveguide with
TWO barriers (undersized guide segments), i.e., with two evanescence
regions[24]. See the text.} \label{fig8}
\end{figure}

\

\begin{figure}[!h]
\begin{center}
 \scalebox{1.6}{\includegraphics{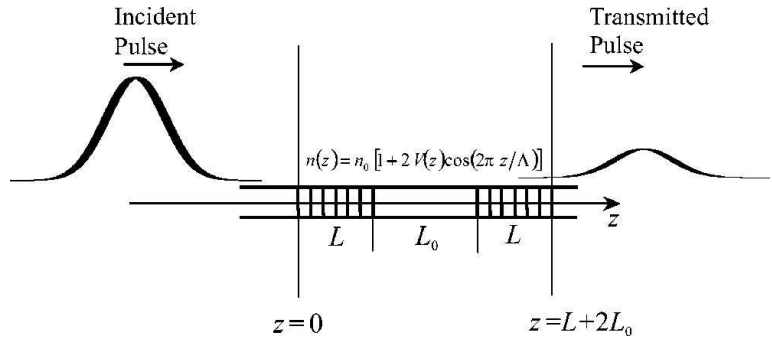}}
\end{center}
\caption{Scheme of tunneling through a rectangular DB photonic structure:
In particular, in ref.[25], as classical barriers there have been
used two gratings in an optical fiber. For the experimental results in the
case of non-resonant tunneling, see
the following figure, Fig.10.} \label{fig9}
\end{figure}

\

\begin{figure}[!h]
\begin{center}
 \scalebox{1.1}{\includegraphics{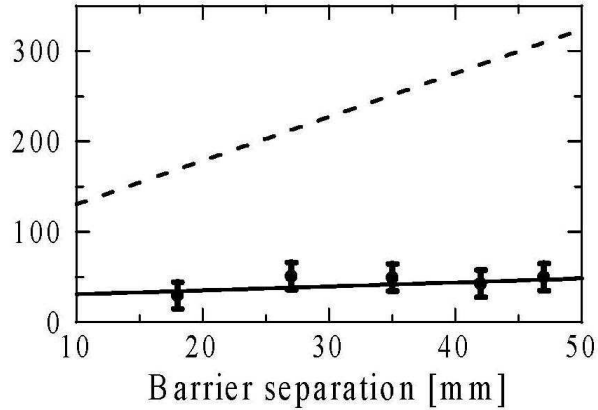}}
\end{center}
\caption{Off-resonance tunnelling time versus barrier separation for
the rectangular symmetric DB FBG structure considered in ref.[25]
(cf. the previous figure, Fig.9). The solid line is the theoretical
prediction based on group delay calculations; {\em the dots are the
experimental points} as obtained by time delay measurements [the dashed curve
is the expected transit time from input to output planes for a
pulse tuned far away from the stopband of the FBGs]. \ The experimental
results in ref.[25] ---as well as the early ones in
refs.[24]--- do confirm the theoretically predicted
independence of the total tunnelling time from the distance between the
two barriers (and, more in general,
the prediction of the ``generalized Hartman Effect[12].} \label{fig10}
\end{figure}

We cannot skip a further topic ---which, being delicate,
should not appear in a brief overview as this one--- since some
experimental contributions to it (like the one performed at Princeton by
J.Wang {\em et al.\/}[23] and published in {\em Nature} on July 20, 2000)
arose a general interest.

Even if in extended relativity all the ordinary causal paradoxes seem to
be solvable[6,3,1] on the basis of the above-seen
Stueckelberg-Feynman-Sudarshan reinterpretation rule, {\em nevertheless}
 ---let us repeat--- one has to remember that (whenever it has to be considered
an object, ${\cal O}$, traveling with Superluminal speed) one can
meet negative contributions to the {\em tunnelling
times\/}[27,12]: and this should not be regarded as unphysical. In
fact, whenever an ``object'' (electromagnetic pulse, particle,...)
${\cal O}$ {\em overcomes} the infinite speed[3,6] with respect to
a certain observer, it will afterwards appear to the same observer
as the ``{\em anti}-object'' $\overline{{\cal O}}$ traveling in
the opposite {\em space} direction[1,3,6].

More precisely, when going on from the lab to a frame ${\cal F}$
moving in the {\em same} direction as the particles or waves
entering the barrier region, the object ${\cal O}$ penetrating and
traveling through the final part of the barrier (with almost
infinite speed[12,21,26,27], like in Figs.3) will appear in the
frame ${\cal F}$ as an anti-object $\overline{{\cal O}}$ crossing
that portion of the barrier {\em in the opposite
space-direction\/}[6,3,1]. In the new frame ${\cal F}$, therefore,
such anti-object $\overline{{\cal O}}$ would yield a {\em
negative} contribution to the tunnelling time: which could even
result, in total, to be negative. For clarifications, see
refs.[28]. What we want to stress here is that the appearance of
such negative times is once more predicted by relativity itself,
on the basis of the ordinary postulates[3,6,21,28]. (In the case
of a non-polarized wave, the wave anti-packet coincides with the
initial wave packet; if a photon is however endowed with helicity
$\lambda =+1$, the anti-photon will bear the opposite helicity
$\lambda =-1$).

From the theoretical point of view, besides refs.[3,6,12,21,27,28], see
refs.[29]. On the (quite interesting!) experimental side, see papers [30].

Let us {\em add} here that, via quantum interference effects
it is possible to obtain dielectrics with refraction indices very rapidly
varying as a function of frequency, also in three-level atomic systems, with
almost complete absence of light absorption (i.e., with quantum induced
transparency)[31]. The group-velocity of a light pulse propagating in
such a medium can decrease to very low values, either positive or negatives,
with {\em no} pulse distortion. It is known that experiments have been
performed both in atomic samples at room temperature, and in Bose-Einstein
condensates, which showed the possibility of reducing the speed of light to
a few meters per second. Similar, but negative group velocities, implying
a propagation with Superluminal speeds thousands of times higher than the
previously mentioned ones, have been recently predicted also in the presence
of such an ``electromagnetically induced transparency'', for light moving in
a rubidium condensate[32], while the corresponding experiments are being
performed (for instance at the ``LENS'' laboratory in Florence).

Finally, let us recall that faster-than-$c$ propagation of
light pulses can be (and was, in same cases) observed also by taking
advantage of anomalous dispersion near an absorbing line, or nonlinear and
linear gain lines ---as already seen---, or nondispersive dielectric media,
or inverted two-level media, as well as of some parametric processes in
nonlinear optics (cf., e.g., G.Kurizki {\em et al.}'s works).

\

{\bf D)} \ {\bf Superluminal Localized Solutions (SLS) to the wave
equations. The ``X-shaped waves"} --- The fourth sector (to leave aside the
others) is not less important. It came into fashion again, when some groups
of scholars in engineering (for sociological reasons, most physicists
had abandoned the field) rediscovered by a series of works that any
wave equation ---to fix the ideas, let us think of the electromagnetic
case--- admit also solutions as much sub-luminal as Super-luminal (besides
the ordinary ---plane, spherical,...--- waves endowed with speed $c/n$).

Let us recall that, starting with the pioneering work by
H.Bateman, it had slowly become known that all wave equations (in
a general sense: scalar, electromagnetic, spinorial,...) admit
wavelet-type solutions with sub-luminal group velocities[33];
namely, soliton-like solutions, even if they are linear equations.
\ Subsequently, also Superluminal solutions started to be written
down.\footnote{This was done in refs.[34] and, independently, in
refs.[35] (in one case just by the mere application of a
Superluminal Lorentz ``transformation"[3,36]).} \ A quite
important feature of some new solutions of these (which attracted
much attention for possible applications) is that they propagate
as {\em localized}, non-diffracting pulses: namely, according to
the Courant and Hilbert's terminology[33], as ``undistorted
progressing waves''; which possess the further property of
``self-reconstructing" themselves after obstacles smaller than
their {\em aperture} size (that is, smaller than the width of the
antenna generating them, enormously larger, in general, than their
wavelenth). It is easy to realize the practical importance, for
instance, of a radio transmission carried out by localized waves,
independently of their being sub- or Super-luminal. But
non-diffractive wave packets can be of use even in theoretical
physics for a reasonable representation of elementary
particles[37]; and so on.

\begin{figure}[!h]
\begin{center}
 \scalebox{1.6}{\includegraphics{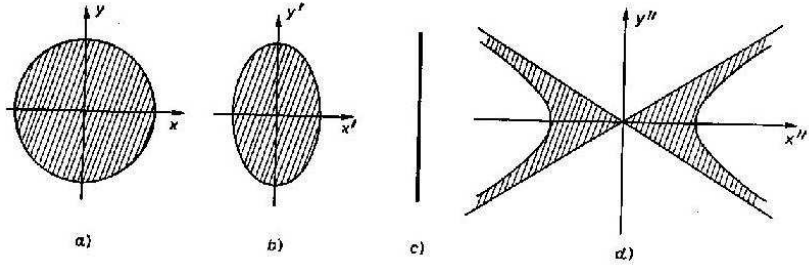}}
\end{center}
\caption{An intrinsically spherical (or pointlike, at the limit)
object appears in the vacuum as an ellipsoid contracted along the
motion direction when endowed with a speed $v<c$. \ By contrast,
if endowed with a speed $V>c$ (even if the $c$-speed barrier
cannot be crossed, neither from the left nor from the right), it
would appear[37] no longer as a particle, but rather as an
``X-shaped" wave[37] traveling rigidly (namely, as occupying the
region delimited by a double cone and a two-sheeted hyperboloid
---or as a double cone, at the limit--, moving Superluminally and
without distortion in the vacuum, or in a homogeneous medium).}
\label{fig11}
\end{figure}

\

\begin{figure}[!h]
\begin{center}
 \scalebox{0.95}{\includegraphics{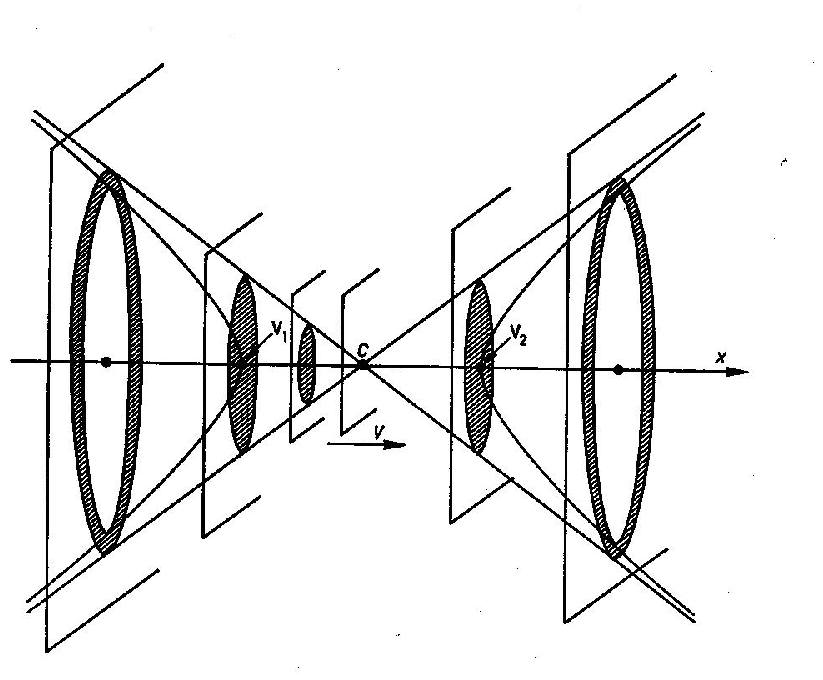}}
\end{center}
\caption{Here we show the intersections of an ``X-shaped wave"[37] with
planes orthogonal to its motion line, according to Extended Relativity``[2-4].
The examination of this figure suggests how to construct a simple dynamic
antenna for generating such localized Superluminal waves (such an antenna
was in fact adopted, independently, by Lu {\em et al.\/}[40] for the production
of such non-diffractive waves).} \label{fig12}
\end{figure}

Within extended relativity since 1980 it had been found
that ---whilst the simplest subluminal object conceivable is a small sphere,
or a point as its limit--- the simplest Superluminal objects turns out to be
instead an ``X-shaped'' wave (see refs.[38], and Figs.11 and 12 of this paper),
or a double cone as its limit, which moreover travels without deforming
---i.e., rigidly--- in a homogeneous medium[3]. \ Analogously, the equipotential
surfaces of the electrostatic field, generated by a tiny charged
sphere at rest, will assume[3,38] the shape represented in Fig.13
when the source is Superluminal. \ For clarifying the connection
existing between what predicted by SR and the localized X-waves
(mathematically and experimentally constructed in recent times, as
we are going to see) let us refer ourselves to a paper appeared in
2004 in Physical Review E, i.e., to the last one of refs.[38],
where the issue of the (X-shaped) field created by a Superluminal
electric charge has been tackled\footnote{At variance with the old
times ---e.g., at the beginning of the seventies our own papers on
similar subjects were always rejected by the ``Phyical
Reviews"---, things have now changed a lot as to superluminal
motions: For instance, the paper of ours quoted as the last one in
Refs.[38], submitted in 2002 to PRL, was diverted to PRE, but was
eventually published therein in 2004, even if dealing with the
X-shaped field generated by a superluminal electric charge! Even
more: at the end of 2007 some authors [S.C.Walker \& W.A.Kuperman,
Phys. Rev. Lett. 99 (2997) 244802] ``imitated" our 1974-PRE
article, without quoting any previous work and not even our 1974
Phys.Rev.E paper!; but it is rather interesting that
Phys.Rev.Lett. published in 2007 an article that ---even if not
original--- was devoted to the field created by a charged
superluminal point-object...}. \ Localized waves do exist, of
course, with any group velocity [the subluminal ones being
ball-like, as expected; see, e.g., the very recent article
``Subluminal wave-bullets: Exact localized subluminal solutions to
the wave equations", Phys.Rev. A77 (2008)033824, by M.Z.Rached and
E.Recami]; but it is not without meaning that the most interesting
localized solutions happened to be just the Superluminal ones, and
with an X-shape. Even more, since from Maxwell equations under
simple hypotheses one goes on to the usual {\em scalar} wave
equation for each electric or magnetic field component, one could
expect the same solutions to exist also in the field of acoustic
waves, and of seismic waves (and of gravitational waves too).

\begin{figure}[!h]
\begin{center}
 \scalebox{0.95}{\includegraphics{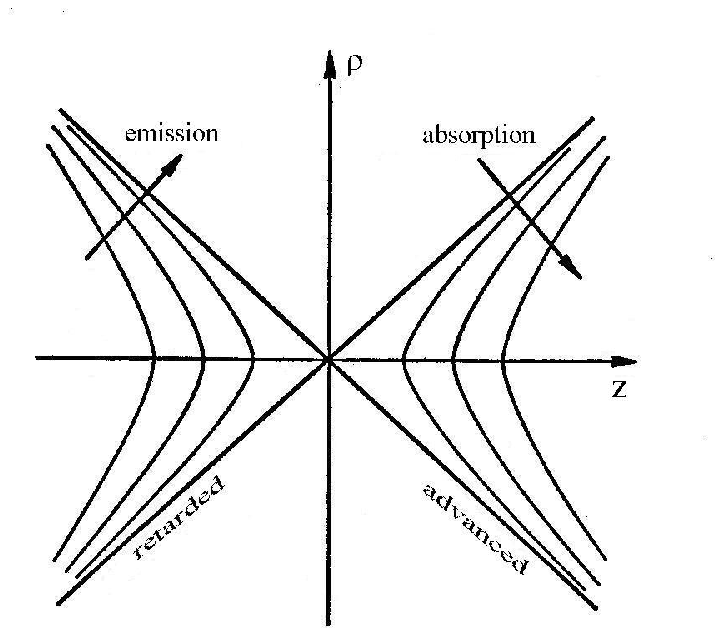}}
\end{center}
\caption{The spherical equipotential surfaces of the electrostatic field
created by a charge at rest get transformed into two-sheeted
rotation-hyperboloids, contained inside an unlimited double-cone, when the
charge travels at Superluminal speed (cf. ref.[3]). This figures shows,
among the others, that a Superluminal charge traveling at constant speed,
in a homogeneous medium like the vacuum, does {\em not} lose energy[3]. \
Let us mention, incidentally, that this double cone has little to do with
the Cherenkov cone. \ The present picture is a reproduction
of our Fig.27, appeared in 1986 at page 80 of ref.[3].} \label{fig13}
\end{figure}

In other words, from the present point of view, it is rather interesting
to note
that, during the last fifteen years, ``X-shaped" waves have been
{\em actually} found as solutions to the Maxwell and to the wave
equations [the form of any wave equations is intrinsically
relativistic, by the way]. \ Actually, such waves (as suitable superpositions
of Bessel beams[39], that is, of simple solutions to the wave equation
already discovered[34] in 1941) were mathematically
constructed for the first time by Lu {\em et al.\/}[40], {\em in acoustics\/}:
and later on by Recami {\em et al.\/}[41] for electromagnetism; and were
then called ``X-waves'' or rather X-shaped waves.
In an elementary Appendix we briefly show
how X-shaped solutions to the wave equation (in particular, the ``classic"
X-wave) can be constructed.

It is more important for us that the X-shaped waves have been
indeed {\em produced in experiments} both with acoustic and with
electromagnetic waves; indeed, X-waves were produced that, in their medium,
travel undistorted with a speed larger than sound, in the first case, and
than light, in the second case. \ In acoustics, the first experiment was
performed by Lu {\em et al.} themselves[42] in 1992, at the Mayo Clinic
(and their papers received the first IEEE 1992 award). In the electromagnetic
case, certainly more intriguing, Superluminal localized X-shaped solutions
were first mathematically constructed (cf., e.g.,
Fig.14) in refs.[41], and
later on experimentally produced by Saari {\em et al.\/}[43] in 1997 at Tartu
by visible light (Fig.15), as announced in their Physical Review Letters
article, and more recently, as we already mentioned, by Ranfagni et al. at
Florence by microwaves[44] (paper appeared in Physical Review letters too,
three years later, in 2000). \ Further experimental activity is in progress;
while in the theoretical sector the activity has been growing so intensely,
that it is not possible to quote here the relevant recent literature; we
might recall, e.g., the papers devoted to building up new
analogous solutions with finite total energy or more suitable for
high frequencies, on one hand, and localized solutions Superluminally propagating
even along a normal waveguide, on the other hand[45,46]; or the attempts
at focusing X-shaped waves, at a certain instant, in a small region[47]. \
But we cannot avoid mentioning that suitable superpositions of Bessel beams
(which can originate also subluminal pulses) can produce
even {\em stationary} intense wave-field: confined within a tiny region
(a static envelope); while the field intensity outside that region is
everywhere negligible[48]; such ``frozen waves" can have
(a patent is pending) very many important applications, for instance as a new
kind of tweezers, and especially ---of course--- in medicine.

\begin{figure}[!h]
\begin{center}
 \scalebox{1.1}{\includegraphics{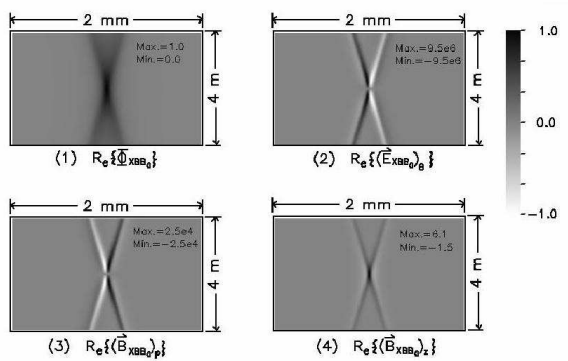}}
\end{center}
\caption{Theoretical prediction of the Superluminal localized ``X-shaped"
waves for the electromagnetic case (from Lu, Greenleaf and Recami[41], and
Recami[41]).} \label{fig14}
\end{figure}

\

\begin{figure}[!h]
\begin{center}
 \scalebox{1.6}{\includegraphics{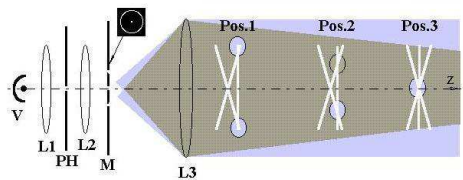}}
\end{center}
\caption{Scheme of the experiment by Saari {\em et al.}, who
announced ({\em Physical Review Letters} of 24 Nov.1997) the
production in optics of the waves depicted in Fig.14: In this
figure one can see what shown by the experiment, i.e., that the
Superluminal ``X-shaped" waves which run after and catch up with
the plane waves (the latter regularly traveling with speed $c$).
An analogous experiment has been performed with microwaves at
Florence by Ranfagni et al. ({\em Physical Review Letters} of 22
May 2000).} \label{fig15}
\end{figure}

Before going on, let us eventually touch the problem of producing an X-shaped
Superluminal wave like the one in Fig.12, but truncated -- of course -- in
space and in time (by the use of a finite, dynamic antenna, radiating
for a finite time): in such a situation, the wave will keep its localization and
Superluminality only along a certain ``depth of field'', decaying abruptly
afterwards[39,41].

We can become convinced about the possibility of realizing it, by
imagining the simple ideal case of a negligibly sized Superluminal
source $S$ endowed with speed $V>c$ in vacuum and emitting
electromagnetic waves $W$ (each one traveling with the invariant
speed $c$). The electromagnetic waves will result to be internally
tangent to an enveloping cone $C$ having $S$ as its vertex, and as
its axis the propagation line $x$ of the source[3].

This is analogous, as we know, to what happens for an airplane that moves in the air with
constant supersonic speed. The
waves $W$ interfere mostly negatively inside the cone $C$, and constructively only
on its surface. We can place a plane detector orthogonally to $x$, and
record magnitude and direction of the $W$ waves that hit on it, as
(cylindrically symmetric) functions of position and of time. It will be
enough, then, to replace the plane detector with a plane antenna which
{\em emits} ---instead of recording--- exactly the same (axially symmetric)
space-time pattern of waves $W$, for constructing a cone-shaped
electromagnetic wave $C$ that will propagate with the Superluminal speed $V$
(of course, without a source any longer at its vertex): even if each wave
$W$ travels\footnote{For further details, see the first of refs.[41].}
with the invariant speed $c$.

Here let us only remark that such localized Superluminal waves
appear to keep their good properties only as long as they are fed
by the waves arriving (with speed $c$) from the antenna: Taking
account of the time needed for fostering such Superluminal pulses
(i.e., for the arrival of the feeding speed-$c$ waves coming from
the aperture). one concludes that these localized Superluminal
waves are probably unable to transmit {\em information} faster
than $c$. However, they don't seem to have anything to do with the
illusory ``scissors effect'', even if the energy feeding them
appears to travel with the speed of light.  In fact, the spot
---endowed, as we know, with Superluminal group-velocity--- is
able to get, for instance, two (tiny) detectors at a distance $L$
to click after a time {\em smaller} than $L/c$). A lot of
discussion is still going on about the possible differences among
group-velocity, signal-velocity and information speed. \ The
interested reader can also check the book {\em Localized Waves}
very recently published by J.Wiley (New York; Jan.2008), ed. by
H.E.H.Figueroa, M.Z.Rached and E.Recami.

As we mentioned above, the existence of all
these X-shaped Superluminal (or ``Super-sonic'') waves seem to constitute at the
moment, together, e.g., with the Superluminality of evanescent waves,
some valuable confirmations of refs.[1], as well as of extended relativity:
a theory[3], let us recall,
based on the ordinary postulates of SR and that consequently does not
appear to violate any of its fundamental principles. It is curious,
moreover,  that one of the first applications of such X-waves (that takes advantage of their
propagation without deformation) is in advanced progress in the field of medicine,
and precisely of ultrasound scanners[49].

Before ending, let us remark that a series of new SLSs to the Maxwell equations, suitable
for arbitrary frequencies and arbitrary bandwidths have been recently constructed
by us: many of them being extremely well localized in the surroundings of
their vertex, and
some of them being endowed with {\em finite} total energy.  Among the others, we have set
forth an infinite family of generalizations of the ``classic" X-shaped wave;
and shown how to deal with the case of a {\em dispersive} medium. Results of this kind
may find application in other fields in which an essential role is played
by a wave-equation.

\

\

{\bf Acknowledgements}\\

The author is deeply indebted, besides to George Sudarshan, to all
the Organizers of this Symposium for their very kind invitation
and hospitality: in particular to Bhamathi Sudarshan, R.M.Walser,
A.Valanju, P.Valanju; and acknowledges the strong cooperation
received along the years by E.C.G.Sudarshan, M.Zamboni Rached,
V.S.Olkhovsky, and H.E.Hern\'andez Figueroa. Thanks are also due,
for stimulating discussions, with G.Marmo, A.Gigli Berzolari,
A.van der Merwe, D.Stauffer; as well as with G.Battistoni,
J.D.Bekenstein, R.Bonifacio, R.Chiao, G.Degli Antoni, F.Fontana,
A.Friberg, A.Hamma, L.Horwitz, M.Ibison, J.Magueijo, P.Milonni,
G.Nimtz, S.Pascazio, P.Pizzochero, A.Ranfagni, P.Saari, A.Shaarawi
and A.Steinberg.

\

\newpage

\cent{{\bf APPENDIX}}

\

\

In this Appendix we want to show how localized superluminal solutions
(SLS) to the wave equations, and in particular the X-shaped ones, can be
mathematically constructed. Here we shall consider, for simplicity, only
the case of a dispersionless medium like vacuum, and of free space (without
boundaries).

\h It is known since more than a century that a particular
axially symmetric solution to the wave equation in vacuum ($n=n_{0}$)
is, in cylindrical coordinates, the function
$ \psi (\rho,z,t) \ug J_0(k_{\rho}\rho) \; \dis{\e^{+ik_z z}
\e^{-i\om t}}$ with $k_{\rho}^2 = \dis{n_0^2\frac{\om^2}{c^2} -
k_z^2}$; \ $k_{\rho}^2 \geq 0 $, where $J_0$ is the zeroth-order
ordinary {\em Bessel function}, $k_z$ and $k_{\rho}$ are the axial and
the transverse wavenumber respectively, $\om$ is the angular
frequency and $c$ the velocity of light. Using the transformation

\begin{equation}
\left\{
\begin{array}{l}
\dis{k_{\rho} = \frac{\om}{c} \, n_{0} \, \sin\,\theta}\\
\\
\dis{k_{z} = \frac{\om}{c} \, n_{0} \, \cos\,\theta}
\end{array}
\right.  \label{transform}
\end{equation}

\noindent
such particular solution $\psi (\rho,z,t)$ can be rewritten in the the
well-known {\em Bessel beam} form:

\begin{equation} \psi (\rho,\ze) \ug J_0(n_{0}\frac{\om}{c} \rho \sin\theta) \;
\dis{\e^{+in_{0}\frac{\om}{c}\ze \cos\theta}} \label{bessel}
\end{equation}

\noindent
where $\ze\equiv z - Vt$ while $V=c/(n_{0}\cos\,\theta)$ is the phase
velocity, quantity $\theta \; (0<\theta<\pi/2)$ being the cone angle of
the Bessel beam.

\h More in general, SLSs (with axial symmetry) to
the wave equation will be the following ones[40,50]:

\begin{equation} \psi (\rho,\ze) \ug
\int_{0}^{\infty}S(\om)J_0\left(\frac{\om}{V} \rho
\sqrt{n_{0}^2\,\frac{V^2}{c^2}-1}\right) \; \dis{\e^{+i
\frac{\om}{V}\ze }} \drm\om  , \label{solgeral}
\end{equation}

where $S(\om)$ is the adopted frequency spectrum.

\h Indeed, such solutions result to be pulses propagating in free space
without distortion and with the Superluminal velocity
$V=c/(n_0\cos\,\theta)$. The most popular spectrum $S(\om)$ is
that one given by $S(\om)=\e^{-a \om}$, which provides the
ordinary (``classic") X-shaped wave

\begin{equation} X\equiv\psi (\rho,\ze) \ug
\frac{V}{\sqrt{(aV-i\ze)^2+\rho^2(n_0^2\frac{V^2}{c^2}-1)}}.
\label{ondaX}
\end{equation}

Because of its non-diffractive properties and its low frequency
spectrum\footnote{Let us emphasize that this spectrum starts
from zero, it being suitable for low frequency applications, and
has the bandwidth $\Delta\om = 1/a$.}, the X-wave is being
particularly applied in fields like acoustics[42].  The
``classic" X-wave is represented in Fig.16. \ As we already said
in the text, infinite series of SLSs can be constructed, more and more
concentrated in the vicinity of their vertex, and corresponding to
any desired frequency and bandwidth.

\begin{figure}[!h]
\begin{center}
 \scalebox{1.6}{\includegraphics{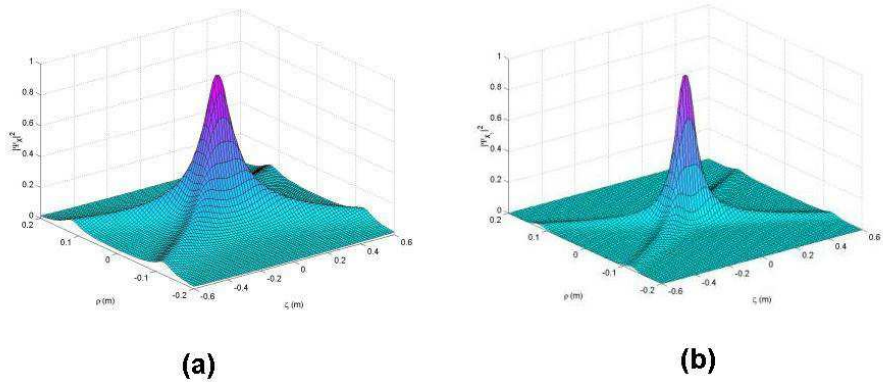}}
\end{center}
\caption{In the first picture it is represented (in arbitrary
units) the square magnitude of the classic $X$-shaped Superluminal
Localized Solution (SLS) to the wave equation, with $V=5\,c$ (and
$a=0.1$): see Refs.[40,41]. \ We have shown elsewhere that
(infinite) families of SLSs however exist, which generalize this
classic $X$-shaped solution:  For instance, the second picture
refers to the ``first derivative" (in the sense specified in
Ref.[45]) of the classic X-wave. \ By increasing the order of the
derivative, the solution gets more and more concentrated in the
surroundings of the Vertex.} \label{fig16}
\end{figure}

\

\newpage

{\bf Figure captions}

\

Fig.1 -- Energy of a free object as a function of its speed.[1-4]\hfill\break

Fig.2 -- Depicting the ``switching rule" (or reinterpretation principle) by
Stueckelberg-Feynman-Sudarshan[1-5]: \ $Q$ will appear as the
antiparticle of P. \ See the text.\hfill\break

Fig.3 -- Behaviour of the average ``penetration time" (in seconds) spent by
a tunnelling wavepacket, as a function of the penetration depth (in
{\aa}angstroms) down a potential barrier (from Olkhovsky {\em et al.}, ref.[12]).
According to the predictions of quantum mechanics, the wavepacket speed
inside the barrier increases in an unlimited way for opaque barriers; and the
total tunnelling time does {\em not} depend on the barrier width[12].\hfill\break

Fig.4 -- Simulation of quantum tunnelling by experiments with
classical evanescent
waves (see the text), which were predicted to be Superluminal also on the
basis of extended relativity[3,4]. The figure shows one of the measurement
results in refs.[15]; that is, the wavepacket average speed while crossing
the evanescent region ( = segment of undersized waveguide, or ``barrier")
as a function of its length. As theoretically predicted[19,12], such an
average speed exceeds $c$ for long enough ``barriers".\hfill\break

Fig.5 -- The delay of a wavepacket crossing a barrier (e.g., a classical
barrier: cf. Fig.4) is due to the initial discontinuity: In
ref.[21]
suitable numerical simulations were therefore performed by considering
an (indefinite) {\em undersized} waveguide, and therefore eliminating
any geometric discontinuity in its cross-section.   \
This figure shows the envelope of the initial signal. \ Inset (a) depicts in
detail the initial part of this signal as a function of time, while inset (b)
depicts the gaussian pulse peak centered at $t = 100$ ns.\hfill\break

Fig.6 -- Envelope of the signal in the previous figure after
having traveled a distance $L = 32.96$ mm through the mentioned
undersized waveguide. \ Inset (a) shows in detail the initial part
(in time) of such arriving signal, while inset (b) shows the peak
of the gaussian pulse that had been initially modulated by
centering it at $t = 100$ ns. \ One can see that its propagation
took {\em zero} time, so that the signal traveled with infinite
speed. \ The numerical simulation has been based on Maxwell
equations only. \ Going on from Fig.5* to this Fig.6* one verifies
that the signal strongly lowered its amplitude: However, the width
of each peak did not change (and this might have some relevance
when thinking of a Morse alphabet ``transmission": see Fig.7* and
the text.\hfill\break

Fig.7 -- As we saw in Figs.5* and 6*, in connection with
(classical, in particular) barriers, a peaked wavepacket suffers a
strong amplitude attenuation while traveling inside a quantum or
classical barrier; its width, however, remains unaffected (this
figure is due to G.Nimtz): Something that, as mentioned in the
previous caption, might have some relevance when thinking of
attempting transmissions by Morse's alphabet.\hfill\break

Fig.8 -- The very interesting experiment along a metallic waveguide with
TWO barriers (undersized guide segments), i.e., with two evanescence
regions[24]. See the text.\hfill\break

Fig.9 -- Scheme of tunneling through a rectangular DB photonic structure:
In particular, in ref.[25], as classical barriers there have been
used two gratings in an optical fiber. For the experimental results in the
case of non-resonant tunneling, see
the following figure, Fig.10*.\hfill\break

Fig.10 -- Off-resonance tunnelling time versus barrier separation for
the rectangular symmetric DB FBG structure considered in ref.[25]
(cf. the previous figure, Fig.9*). The solid line is the theoretical
prediction based on group delay calculations; {\em the dots are the
experimental points} as obtained by time delay measurements [the dashed curve
is the expected transit time from input to output planes for a
pulse tuned far away from the stopband of the FBGs]. \ The experimental
results in ref.[25] ---as well as the early ones in
refs.[24]--- do confirm the theoretically predicted
independence of the total tunnelling time from the distance between the
two barriers (and, more in general,
the prediction of the ``generalized Hartman Effect[12].\hfill\break

Fig.11 -- An intrinsically spherical (or pointlike, at the limit)
object appears in the vacuum as an ellipsoid contracted along the
motion direction when endowed with a speed $v<c$. \ By contrast,
if endowed with a speed $V>c$ (even if the $c$-speed barrier
cannot be crossed, neither from the left nor from the right), it
would appear[37] no longer as a particle, but rather as an
``X-shaped" wave[37] traveling rigidly (namely, as occupying the
region delimited by a double cone and a two-sheeted hyperboloid
---or as a double cone, at the limit--, moving Superluminally and
without distortion in the vacuum, or in a homogeneous
medium).\hfill\break

Fig.12 -- Here we show the intersections of an ``X-shaped wave"[37] with
planes orthogonal to its motion line, according to Extended Relativity``[2-4].
The examination of this figure suggests how to construct a simple dynamic
antenna for generating such localized Superluminal waves (such an antenna
was in fact adopted, independently, by Lu {\em et al.\/}[40] for the production
of such non-diffractive waves).\hfill\break

Fig.13 -- The spherical equipotential surfaces of the electrostatic field
created by a charge at rest get transformed into two-sheeted
rotation-hyperboloids, contained inside an unlimited double-cone, when the
charge travels at Superluminal speed (cf. ref.[3]). This figures shows,
among the others, that a Superluminal charge traveling at constant speed,
in a homogeneous medium like the vacuum, does {\em not} lose energy[3]. \
Let us mention, incidentally, that this double cone has little to do with
the Cherenkov cone. \ The present picture is a reproduction
of our Fig.27, appeared in 1986 at page 80 of ref.[3].\hfill\break

Fig.14 -- Theoretical prediction of the Superluminal localized ``X-shaped"
waves for the electromagnetic case (from Lu, Greenleaf and Recami[41], and
Recami[41]).\hfill\break

Fig.15 -- Scheme of the experiment by Saari {\em et al.}, who
announced ({\em Physical Review Letters} of 24 Nov.1997) the
production in optics of the waves depicted in Fig.8: In this
figure one can see what shown by the experiment, i.e., that the
Superluminal ``X-shaped" waves which run after and catch up with
the plane waves (the latter regularly traveling with speed $c$).
An analogous experiment has been performed with microwaves at
Florence by Ranfagni et al. ({\em Physical Review Letters} of 22
May 2000).\hfill\break

Fig.16 -- In the first picture it is represented (in arbitrary units) the square
magnitude of the classic $X$-shaped Superluminal Localized Solution (SLS)
to the wave equation, with $V=5\,c$ (and $a=0.1$): see Refs.[40,41]. \
We have shown elsewhere that (infinite) families of SLSs however exist,
which generalize this classic $X$-shaped solution:  For instance, the second
picture refers to the ``first derivative" (in the sense specified in
Ref.[45]) of the classic X-wave.\hfill\break

\newpage

{\bf Bibliography:}\hfill\break

[1] O.M.Bilaniuk, V.K.Deshpande and E.C.G.Sudarshan, {\em Am. J.
Phys.}, vol.30, 718 (1962); \ E.C.G.Sudarshan: J. Math. Phys. (NY), vol.4,
1029 (1963); Ark. Fiz. 39, 585 (1969); Proc. Indian Acad. Sci., vo.69, 133
(1969); in {\em Symposia on Theoretical Physics and Mathematics}, vol.10, 129
(New York, NY, 1970); Phys. Rev., vol.D1, 2473 (1970); in {\em Proceedings of the
VIII Nobel Symposium}, ed. by N.Swartholm (New York, NY, 1970), p.385;
Reports NYO-3399-191/SU-1206-191 (Syracuse University, 1969), ORO-3992-5
(UT, Austin, 1970), CPT-81/AEC-30 (UT, Austin, 1970) and CPT-166 (UT, Austin,
1972); \ M.E.Arons and E.C.G.Sudarshan, Phys. Rev., vol.173, 1622 (1968); \
J.Dhar and E.C.G.Sudarshan, Phys. Rev., vol.174, 1808 (1968); \ O.M.B.Bilaniuk
and E.C.G.Sudarshan, Nature, vol.223, 386 (1969); Phys. Today, vol.22, 43
(1969); \ E.C.G.Sudarshan and N.Mukunda, Phys. Rev., vol.D1, 571 (1970); \
A.M.Gleeson and E.C.G.Sudarshan, Phys. Rev. D1, 474 (1970); \
A.M.Gleeson, M.G.Gundzig, E.C.G.Sudarshan and A.Pagnamenta: Phys. Rev.,
vol.A6, 807 (1972); {\em Fields Quanta,} vol.2, 175 (1972); \
E.C.G.Sudarshan and J.V.Narlikar, Montly Notices of Royal Astron. Soc.,
vol.175, 105 (1976); \ E.C.G.Sudarshan, in {Tachyons, Monopoles, and
Related Topics}, ed. by E.Recami (North-Holland; Amsterdam, 1978),
pp.43-46.\hfill\break

[2] For an early review, see E.Recami and R.Mignani, {\em Rivista N. Cim.},
vol.4, 209-290, E398 (1974), and refs. therein. \ Cf. also E.Recami (editor),
{\em Tachyons,
Monopoles, and Related Topics} (North-Holland; Amsterdam, 1978); and
T.Alv\"ager and M.N.Kreisler, {\em Phys. Rev.}, vol.171, 1357 (1968) and
refs. therein.\hfill\break

[3] E.Recami, {\em Rivista N. Cim.}, vol.9(6), 1$\div$178
(1986), issue no.6 [178 printed pages], and refs. therein.\hfill\break

[4] See O.M.Bilaniuk, V.K.Deshpande and E.C.G.Sudarshan, ref.[1]. \ Cf. also,
e.g., E.Recami, in {\em Annuario '73, Enciclopedia EST}, ed. by
E.Macorini (Mondadori; Milano, 1973), pp.85-94; \ and \ {\em Nuovo Saggiatore},
vol.2(3), 20-29 (1986), issue no.3.\hfill\break

[5] E.Recami, in {\em I Concetti della Fisica}, ed. by F.Pollini and
G.Tarozzi (Acc. Naz. Sc. Lett. Arti; Modena, 1993), pp.125-138; \ E.Recami and
W.A.Rodrigues, ``Antiparticles from Special Relativity", {\em Found. Physics},
vol.12, 709-718 (1982); \, vol.13, E533 (1983).\hfill\break

[6] E.Recami, {\em Found. Physics}, vol.17, 239-296 (1987). \ See also
{\em Lett. Nuovo Cimento}, vol.44, 587-593 (1985); \ P.Caldirola and
E.Recami, in {\em Italian Studies in the Philosophy of Science}, ed. by
M.Dalla Chiara (Reidel; Boston, 1980), pp.249-298; \ A.M.Shaarawi
and I.M.Besieris, {\em J. Phys. A}, vol.33, 7255-7263 (2000).\hfill\break

[7] Cf., e.g., M.Baldo Ceolin, ``Review of neutrino physics", invited talk at the
{\em XXIII Int. Symp. on Multiparticle Dynamics (Aspen, CO; Sept.1993)}; \
E.W.Otten, {\em Nucl. Phys. News}, vol.5, 11 (1995). \ From the theoretical
point of view, see E.Giannetto, G.D.Maccarrone, R.Mignani and E.Recami,
{\em Phys. Lett. B}, vol.178, 115-120 (1986) and refs. therein; \ S.Giani,
``Experimental evidence of superluminal velocities in astrophysics and proposed
experiments", CP458, in {\em Space Technology and Applications International
Forum 1999}, ed. by M.S.El-Genk (A.I.P.; Melville, 1999), pp.881-888; \
R.G.Cawley, Lett. Nuovo Cim., vol.3, 523 (1972).\hfill\break

[8] See, e.g., J.A.Zensus and T.J.Pearson (editors), {\em Superluminal Radio
Sources} (Cambridge Univ.Press; Cambridge, UK, 1987).\hfill\break

[9] I.F.Mirabel and L.F.Rodriguez, ``A superluminal source in the Galaxy",
{\em Nature}, vol.371, 46 (1994) [with an editorial comment, ``A galactic speed record",
by G.Gisler, at page 18 of the same issue]; \ S.J.Tingay {\em et al.},
``Relativistic motion in a nearby bright X-ray source", {\em Nature},
vol.374, 141 (1995).\hfill\break

[10] M.J.Rees, {\em Nature}, vol.211, 46 (1966); \ A.Cavaliere, P.Morrison and L.Sartori,
{\em Science}, vol.173, 525 (1971).\hfill\break

[11] E.Recami, A.Castellino, G.D.Maccarrone and M.Rodon\`o, ``Considerations
about the apparent Superluminal expansions observed in astrophysics", {\em Nuovo
Cimento B}, vol.93, 119 (1986). \ Cf. also R.Mignani and E.Recami, {\em Gen. Relat. Grav.},
vol.5, 615 (1974).\hfill\break

[12] V.S.Olkhovsky, E.Recami and J.Jakiel, ``Unified time analysis of photon and
particle tunnelling", {\em Phys. Reports}, vol.398, pp.133-178 (2004);
V.S.Olkhovsky and E.Recami, {\em Phys. Reports}, vol.214, 339 (1992), and refs.
therein, in particular T.E.Hartman, {\em J. Appl. Phys.}, vol.33, 3427 (1962); \
L.A.MacColl, {Phys. Rev.}, vol.40, 621 (1932). \
See also V.S.Olkhovsky, E.Recami, F.Raciti and A.K.Zaichenko,
{\em J. de Phys.-I}, vol.5, 1351-1365 (1995); \
G.Privitera, E.Recami, G.Salesi and V.S.Olkhovsky: ``Tunnelling Times:
An Elementary Introduction",
{\em Rivista Nuovo Cim.}, vol.26 (2003), monographic issue no.4.

[13] See, e.g., R.Y.Chiao, P.G.Kwiat and A.M.Steinberg, {\em Physica B}, vol.175,
257 (1991); \ A.Ranfagni, D.Mugnai, P.Fabeni and G.P.Pazzi, {\em Appl. Phys. Lett.},
vol.58, 774 (1991); \ Th.Martin and R.Landauer, {\em Phys. Rev. A}, vol.45,
2611 (1992); \ Y.Japha and G.Kurizki, {\em Phys. Rev. A},
vol.53, 586 (1996). \ Cf. also G.Kurizki, A.E.Kozhekin and A.G.Kofman,
{\em Europhys. Lett.}, vol.42, 499 (1998); \
G.Kurizki, A.E.Kozhekin, A.G.Kofman and M.Blaauboer, paper delivered at the
VII Seminar on Quantum Optics, Raubichi, BELARUS (May, 1998).\hfill\break

[14] E.Recami, F.Fontana and R.Garavaglia, {\em Int. J. Mod. Phys. A},
vol.15, 2793 (2000), and refs. therein.\hfill\break

[15] G.Nimtz and A.Enders, {\em J. de Physique-I}, vol.2, 1693 (1992);
vol.3, 1089 (1993); \, vol.4, 1379 (1994); \ {\em Phys. Rev. E},
vol.48, 632 (1993); \ H.M.Brodowsky, W.Heitmann and G.Nimtz,
{\em J. de Physique-I}, vol.4, 565 (1994); \ {\em Phys. Lett. A},
vol.222, 125 (1996); \, vol.196, 154 (1994). \ For a review, see, e.g.,
G.Nimtz and W.Heitmann, {\em Prog. Quant. Electr.}, vol.21, 81 (1997).\hfill\break

[16] A.M.Steinberg, P.G.Kwiat and R.Y.Chiao, {\em Phys. Rev. Lett.},
vol.71, 708 (1993), and refs. therein; \ {\em Scient. Am.}, vol.269(2),
38 (1993), issue no.2. \ For a review, see, e.g., R.Y.Chiao and A.M.Steinberg,
in {\em Progress in Optics}, ed. by E.Wolf (Elsevier; Amsterdam, 1997), p.345.
 \ Cf. also Y.Japha and G.Kurizki, {\em Phys. Rev. A},
vol.53, 586 (1996).\hfill\break

[17] A.Ranfagni, P.Fabeni, G.P.Pazzi and D.Mugnai, {\em Phys. Rev. E},
vol.48, 1453 (1993); \ Ch.Spielmann, R.Szipocs, A.Stingl and F.Krausz,
{\em Phys. Rev. Lett.}, vol.73, 2308 (1994), \ Ph.Balcou and L.Dutriaux,
{\em Phys. Rev. Lett.}, vol.78, 851 (1997); \ V.Laude and P.Tournois,
{\em J. Opt. Soc. Am. B}, vol.16, 194 (1999).\hfill\break

[18] {\em Scientific American} (Aug. 1993); \ {\em Nature} (Oct.21, 1993); \ {\em New
Scientist} (Apr. 1995); \ {\em Newsweek} (19 June 1995).\hfill\break

[19] Ref.[3], p.158 and pp.116-117. \ Cf. also D.Mugnai, A.Ranfagni,
R.Ruggeri, A.Agresti and E.Recami, {\em Phys. Lett. A}, vol.209, 227
(1995).\hfill\break

[20] H.M.Brodowsky, W.Heitmann and G.Nimtz, {\em Phys. Lett. A},
vol.222, 125 (1996).\hfill\break

[21] A.P.L.Barbero, H.E.H.Figueroa, and E.Recami, ``On the propagation
speed of evanescent modes", {\em Phys. Rev. E}, vol.62, 8628 (2000),
and refs. therein.

[22] Cf., e.g., P.W.Milonni, {J. Phys. B}, vol.35 (2002) R31-R56; \ 
G.Nimtz and A.Haibel, {Ann. der Phys.}, vol.11 (2002) 163-171; \
R.W.Ziolkowski, {Phys. Rev. E}, vol.63 (2001) no.046604; \
A.M.Shaarawi and I.M.Besieris, {J. Phys. A}, vol.33 (2000) 7227-7254;
7255-7263; \ E.Recami, F.Fontana and R.Garavaglia, {\em Int. J. Mod. Phys. A},
vol.15, 2793 (2000), and refs. therein.\hfill\break

[23] L.J.Wang, A.Kuzmich and A.Dogariu, {\em Nature}, vol.406 (2000) 277.\hfill\break

[24] G.Nimtz, A.Enders and H.Spieker, in {\em Wave and Particle in Light and
Matter}, ed. by A.van der Merwe and A.Garuccio (Plenum; New York, 1993); \ {\em J.
de Physique-I}, vol.4, 565 (1994). \ See also A.Enders and G.Nimtz,
{\em Phys. Rev. B}, vol.47, 9605 (1993).\hfill\break

[25] S.Longhi, P.Laporta, M.Belmonte and E.Recami, {\em Phys. Rev. E}, vol.65
(2002) no.046610.\hfill\break

[26] V.S.Olkhovsky, E.Recami and G.Salesi, {\em Europhysics Letters} 57 (2002)
879-884; \ ``Tunneling through two successive barriers and the Hartman
(Superluminal) effect", e-print quant-ph/0002022; \ Y.Aharonov, N.Erez and
B.Reznik, {\em Phys. Rev. A}, vol.65 (2002) no.052124. \
See also E.Recami: ``Superluminal tunneling through successive barriers: Does
QM predict infinite group-velocities?", {\em Journal of Modern Optics},
vol.51, pp.913-923 (2004); \ V.S.Olkhovsky, E.Recami and A.K.Zaichenko:
``Resonant and non-resonant tunneling through a double barrier",
{\em Europhysics Letters}, vol.70, 712-718 (2005).\hfill\break

[27] V.S.Olkhovsky, E.Recami and G.Salesi, refs.[26]; \ S.Esposito,
{\em Phys. Rev. E}, vol.67 (2003) no.016609.\hfill\break

[28] V.S.Olkhovsky, E.Recami, F.Raciti and A.K.Zaichenko, ref.[12],
page 1361 and refs. therein. \ See also refs.[3,6], and E.Recami,
F.Fontana and R.Garavaglia, ref.[14], page 2807 and refs.
therein.\hfill\break

[29] R.Y.Chiao, A.E.Kozhekin A.E., and G.Kurizki, {\em Phys. Rev. Lett.},
vol.77, 1254 (1996); \ E.L.Bolda {\em et al.}, {Phys. Rev. A}, vol.48, 3890
(1993); \ C.G.B.Garret and D.E.McCumber, {\em Phys. Rev. A},
vol.1, 305 (1970).\hfill\break

[30] L.J.Wang, A.Kuzmich and A.Dogariu, {\em Nature}, vol.406 (2000) 277; \
M.W.Mitchell and R.Y.Chiao, {\em Phys. Lett. A}, vol.230 (1997) 133-138. \
See also S.Chu and Wong W., {\em Phys. Rev. Lett.}, vol.48 (1982) 738; \
B.Segard and B.Macke, {\em Phys. Lett. A}, vol.109 (1985) 213-216; B.Macke
{\em et al.}, {\em J. Physique}, vol.48 (1987) 797-808; \ G.Nimtz, {\em Europ.
Phys. J., B} (to appear as a Rapid Note).\hfill\break

[31] G.Alzetta, A.Gozzini, L.Moi and G.Orriols, {\em Nuovo Cimento B},
vol.36, 5 (1976).\hfill\break

[32] M.Artoni, G.C.La Rocca, F.S.Cataliotti and F.Bassani, {\em Phys. Rev. A} (in
press).\hfill\break

[33] H.Bateman, {\em Electrical and Optical Wave Motion} (Cambridge
Univ.Press; Cambridge, 1915); \ R.Courant and D.Hilbert, {\em Methods of
Mathematical Physics} (J.Wiley; New York, 1966), vol.2, p.760; \
J.N.Brittingham, {\em J. Appl. Phys.}, vol.54, 1179 (1983); \ R.W.Ziolkowski,
{\em J. Math. Phys.}, vol.26, 861 (1985); \ J.Durnin, J.J.Miceli and
J.H.Eberly, {\em Phys. Rev. Lett.}, vol.58, 1499 (1987); \ A.O.Barut {\em et al.},
{\em Phys. Lett. A}, vol.143, 349
(1990); \ {\em Found. Phys. Lett.}, vol.3, 303 (1990); \
{\em Found. Phys.}, vol.22, 1267 (1992); \ P.Hillion, {\em Acta
Applicandae Matematicae}, vol.30 (1993) 35.\hfill\break

[34] J.A.Stratton, {\em Electromagnetic Theory} (McGraw-Hill; New York,
1941), p.356; \ A.O.Barut {\em et al.}, {\em Phys. Lett. A}, vol.180, 5 (1993); \,
vol.189, 277 (1994). \ For review-articles about Localized Superluminal
Waves, see E.Recami, M.Z.Rached, K.Z.n\'obrega, C.A.Dartora and
H.E.H.Figueroa: ``On the localized superluminal solutions to the Maxwell
equations", {\em IEEE Journal of Selected Topics in Quantum Electronics},
vol.9(1) 59-73 (2003); and the book {\em Localized Waves}, ed. by
H.E.H.Figueroa, M.Z.Rached and E.Recami (J.Wiley; New York,
Jan.2008).\hfill\break

[35] R.Donnelly and R.W.Ziolkowski, {\em Proc. Roy. Soc. London A}, vol.440, 541 (1993); \
I.M.Besieris, A.M.Shaarawi and R.W.Ziolkowski, {\em J. Math. Phys.},
vol.30, 1254 (1989); \ S.Esposito, {\em Phys. Lett. A}, vol.225, 203
(1997); \ J.Vaz and W.A.Rodrigues, {\em Adv. Appl. Cliff. Alg.},
vol.S-7, 457 (1997).\hfill\break

[36] See also E.Recami and W.A.Rodrigues Jr., ``A model theory for tachyons
in two dimensions", in {\em Gravitational Radiation and Relativity}, ed. by
J.Weber and T.M.Karade (World Scient.; Singapore, 1985), pp.151-203, and
refs. therein.\hfill\break

[37] A.M.Shaarawi, I.M.Besieris and R.W.Ziolkowski, {\em J. Math. Phys.},
vol.31, 2511 (1990), Sect.VI; \ {\em Nucl Phys. (Proc.Suppl.) B},
vol.6, 255 (1989); \ {\em Phys. Lett. A}, vol.188, 218 (1994). \ See also
V.K.Ignatovich, {\em Found. Phys.}, vol.8, 565 (1978); and
A.O.Barut, {\em Phys. Lett. A}, vol.171, 1 (1992); \, vol.189, 277 (1994); \
{\em Ann. Foundation L. de Broglie}, Jan.1994; and ``Quantum theory of single
events, Localized de Broglie--wavelets, Schroedinger waves and classical trajectories",
preprint IC/90/99 (ICTP; Trieste, 1990).\hfill\break

[38] A.O.Barut, G.D.Maccarrone and E.Recami, {\em Nuovo Cimento A},
vol.71, 509 (1982); \ P.Caldirola, G.D.Maccarrone and E.Recami,
{\em Lett. Nuovo Cim.}, vol.29, 241 (1980); \ E.Recami and G.D.Maccarrone,
{\em Lett. Nuovo Cim.}, vol.28, 151 (1980). \ See also E.Recami, refs.[3,4,41],
and E.Recami, M.Z.Rached and C.A.Dartora: ``The X-shaped, localized
field generated by a Superluminal electric charge", {\em Phys. Rev. E},
vol.69 (2004) no.027602.\hfill\break

[39] J.Durnin, J.J.Miceli and J.H.Eberly, {\em Phys. Rev. Lett.}, vol.58 (1987)
1499; \ {\em Opt. Lett.}, vol.13, 79 (1988).\hfill\break

[40] J.-y.Lu and J.F.Greenleaf, {\em IEEE Trans. Ultrason. Ferroelectr. Freq.
Control}, vol.39, 19 (1992).\hfill\break

[41] E.Recami, {\em Physica A}, vol.252, 586 (1998); \ J.-y.Lu,
J.F.Greenleaf and E.Recami, ``Limited diffraction solutions to
Maxwell (and Schroedinger) equations'' [Lanl Archives \#
physics/9610012], Report INFN/FM--96/01 (I.N.F.N.; Frascati,
Oct.1996). \ See also R.W.Ziolkowski, I.M.Besieris and
A.M.Shaarawi, {\em J. Opt. Soc. Am., A}, vol.10, 75 (1993); {\em
J. Phys. A}, vol.33, 7227-7254 (2000); \ A.T.Friberg, A.Vasara and
J.Turunen: {\em Phys. Rev. A}, vol.43, 7079 (1991); \ and the whole
book on {\em Localized Waves} quoted under Ref.[34].\hfill\break

[42] J.-y.Lu and J.F.Greenleaf, {\em IEEE Trans. Ultrason. Ferroelectr. Freq.
Control}, vol.39, 441 (1992): In this case the wave speed is larger than the
{\em sound} speed in the considered medium.\hfill\break

[43] P.Saari and K.Reivelt, ``Evidence of X-shaped propagation-invariant
localized light waves", {\em Phys. Rev. Lett.}, vol.79, 4135-4138
(1997).\hfill\break

[44] D.Mugnai, A.Ranfagni and R.Ruggeri, {\em Phys. Rev. Lett.},
vol.84, 4830 (2000).\hfill\break

[45] M.Z.Rached, E.Recami and H.E.H.Figueroa, ``New localized
Superluminal solutions to the wave equations with finite total
energies and arbitrary frequencies", {\em European Physical
Journal D}, vol.21, pp.217-228 (2002); M.Z.Rached, K.Z.N\'obrega,
H.E.H.Figueroa, and E.Recami: "Localized Superluminal solutions to
the wave equation in (vacuum or) dispersive media, for arbitrary
frequencies and with adjustable bandwidth", {\em Optics
Communications}, vol.226 15-23 (2003); \ M.Z.Rached, E.Recami and
F.Fontana, ``Superluminal localized solutions to Maxwell equations
propagating along a waveguide: The finite-energy case", {\em
Physical Review E}, vol.67 (2003) no.036620. \ See also the first
two (introductory) Chapters in the volume {\em Localized Waves}
quoted in Ref.[34]; \ and M.Z.Rached and E.Recami, ``Subluminal wabe
ullets: Exact localized subluminal solutions to the wave equations",
{\em Physical Review} A77 (2008) 033824.\hfill\break

[46] M.Z.Rached, E.Recami and F.Fontana, ``Localized Superluminal solutions
to Maxwell equations propagating along a normal-sized waveguide'',
{\em Phys. Rev. E}, vol.64 (2001) no.066603; \
M.Z.Rached, K.Z.Nobrega, E.Recami and H.E.H.Figueroa, ``Superluminal
X-shaped beams propagating without distortion along a co-axial guide",
{\em Physical Review E}, vol.66 (2002) no.046617.; \ I.M.Besieris,
M.A.Rahman,
A.Shaarawi and A.Chatzipetros, {\em Progress in Electromagnetic Research
(PIER)}, vol.19, 1-48 (1998).\hfill\break

[47]  M.Z.Rached, A.M.Shaarawi and E.Recami: ``Focused X-shaped pulses",
{\em Journal of the Optical Society of America A}, vol.21, pp.1564-1574
(2004), and refs. therein.\hfill\break

[48] M.Z.Rached: ``Stationary optical wave-fields with arbitrary longitudinal
shape", {\em Optics Express}, vol.12, pp.4001-4006 (2004).\hfill\break

[49] J.-y.Lu, H.-h.Zou and J.F.Greenleaf, {\em Ultrasound in Medicine and
Biology}, vol.20, 403 (1994); \ {\em Ultrasonic Imaging}, vol.15, 134
(1993).\hfill\break

[50] H.S\~{o}najalg, P.Saari, ``Suppression of
temporal spread of ultrashort pulses in dispersive media by
Bessel beam generators'', {\em Opt. Letters}, vol.21,
pp.1162-1164, August 1996.

\end{document}